\documentclass{aa}  

\usepackage{graphicx}
\usepackage{subcaption}
\usepackage{txfonts}
\usepackage{newtxtext,newtxmath}
\usepackage{siunitx}
\usepackage[colorlinks=true, linkcolor=blue, citecolor=blue, filecolor=blue,urlcolor=blue]{hyperref}
\usepackage[capitalize]{cleveref}

\AtBeginEnvironment{appendices}{\crefalias{section}{appendix}}

\begin{document}

   \title{Disentangling the Galactic binary zoo: Machine learning classification of stellar remnant binaries in LISA data}

   \titlerunning{Machine learning classification of Galactic stellar remnant binaries in LISA data}

   \author{Irwin K. C. Tay
          \inst{1}
          \and
          Valeriya Korol\inst{2,1}
          \and Thibault Lechien\inst{1}
          }

   \institute{Max-Planck-Institut f{\"u}r Astrophysik, Karl-Schwarzschild-Stra{\ss}e 1, 85748 Garching, Germany
   \and
            SRON Space Research Organisation Netherlands, Niels Bohrweg 4, 2333 CA Leiden, the Netherlands \\
              \email{v.korol@sron.nl}
             }


 
\abstract{
The Laser Interferometer Space Antenna (LISA) will open a new observational window in the millihertz gravitational-wave band, enabling the detection of tens of thousands of compact stellar remnant binaries across the Milky Way. Most of LISA’s sources will be double white dwarf (WDWD) systems, while neutron star–white dwarf (NSWD) binaries and higher-mass systems will be orders of magnitude rarer but of significant astrophysical interest. Disentangling these populations is challenging due to the strong overlap in their gravitational-wave features. In this work, we investigate the use of machine-learning techniques to classify LISA-detectable binaries based solely on LISA observables.

Using mock catalogues of Galactic binaries constructed from population-synthesis studies, we evaluate a range of machine-learning classifiers. We find that ensemble-based methods—particularly gradient-boosting algorithms such as XGBoost—deliver the best performance on our highly imbalanced dataset. WDWD systems are identified with a recall of $\sim 99\%$, reflecting their dominant presence, and high-mass binaries are also classified with high recall ($\ge 85\%$). In contrast, NSWD systems remain the most challenging population to distinguish: their features overlap strongly with those of WDWD binaries, making them particularly prone to misclassification. Despite this, XGBoost correctly identifies 85.6\% of NSWD systems in our simulated LISA detections, outperforming simple statistical approaches based on kernel density estimation.
We further demonstrate that machine-learning classification can effectively support the interpretation of LISA data, enabling the identification of eccentric binaries and extremely rare subclasses.
}

   \maketitle
%

\section{Introduction}
The Laser Interferometer Space Antenna (LISA) is one of the European Space Agency’s flagship missions, scheduled for launch in the 2030s \citep{Redbook}. 
It will open a new observational window in the gravitational-wave spectrum, between 0.1~mHz and 1~Hz, providing access to a new set of gravitational-wave sources: from compact stellar remnant binaries in our own Galaxy, to mergers of massive black holes across cosmic time, and to potential cosmological signals from the primordial Universe \citep[for a review see][]{FundPhysicsWP, CosmoWP, AstroWP}. 

In this study, we consider Galactic gravitational-wave sources accessible to LISA—binary systems composed of white dwarfs, neutron stars, and black holes with orbital periods of less than a few hours. Among these, double white dwarf (WDWD) binaries are expected to be the most numerous by far \citep[e.g.][]{Nelemans2001}. Theoretical studies predict that LISA will individually resolve approximately $\mathcal{O}(10^4)$ WDWD binaries \citep[e.g.][]{Korol2017, lam19, Thiele2023, li2023, Abinaya2025}. This is mainly due to two factors. First, the initial mass function favours the formation of low-mass stars, which later evolve into white dwarfs. Second, low-mass stars live much longer than their massive counterparts and therefore accumulate in the Galaxy over time. Comparatively, all other types of binaries are expected to be orders of magnitude less numerous: binaries composed of a neutron star and a white dwarf (NSWD) are expected to yield $\mathcal{O}(10^2)$ detections \citep[e.g.][]{bre20, chen2020, Korol2024Ia}, while binaries containing two neutron stars (NSNS) or black holes (BHNS and BHBH) are predicted to be even rarer, with only $\mathcal{O}(10)$ expected detections for each of these classes \citep[e.g.][]{lam18,lau2020,Wagg2022,tang24}. These latter systems are the products of massive stellar evolution and, owing to their short lifespans and the supernova kicks that can disrupt them, they tend to disappear from the Galaxy over time \citep[e.g.][]{Wagg2025}.

These detections represent only the tip of the iceberg of the underlying Galactic population in the LISA frequency band. The diverse binary populations will together produce a complex gravitational-wave signal in the LISA band, composed of tens of millions of overlapping sources. Extracting the thousands of individually `loud' signals mentioned above from the overall Galactic `hum' represents one of the central challenges of LISA data analysis \citep[e.g.][]{Nissanke2012,Karnesis2021,2023A&A...678A.123L}. Addressing this challenge is the first step in LISA’s global fit analysis, which involves the simultaneous modelling and subtraction of multiple source types from the data stream \citep[e.g.][]{Littenberg2023, Strub2024, Katz2025, Deng2025}. 

Assuming that the LISA global fit successfully extracts and catalogues $\mathcal{O}(10^4)$ individual Galactic gravitational-wave sources, a natural next question arises: given the overwhelming number of WDWD binaries, how easily, if at all, can the different types of binaries be distinguished from each other? This is a non-trivial task, since in the LISA band most binaries appear as continuous, quasi-monochromatic sources. In the low-frequency regime ($f \lesssim 1$–$2\,\mathrm{mHz}$), their gravitational-wave signal is effectively monochromatic, showing no measurable frequency derivative (the so-called chirp); in this case only the amplitude and frequency can be determined from the LISA data alone. Without a chirp measurement, the chirp mass—the specific combination of the component masses that sets the frequency evolution—is degenerate with distance, making it virtually impossible to distinguish between, for example, a nearby WDWD binary and a more distant BHBH binary \citep[e.g.][]{Sesana2020}. At higher frequencies ($f \gtrsim 1-3\,\mathrm{mHz}$), the frequency derivative becomes measurable, allowing one to infer bounds on binary component masses via the chirp mass measurement \citep[e.g.][]{Korol2024Ia}. 
Recent studies by \citet{Moore2024} and \citet{Middleton2025} have further shown that, if the binary is eccentric, individual component masses can be constrained, at least in some cases. In eccentric systems, the orbit will undergo periastron precession, which introduces an additional modulation in the gravitational-wave signal. The associated precession frequency depends on the total mass of the binary, and if it can be measured together with the chirp mass, the degeneracy is broken and the individual component masses can be disentangled \citep{Seto2001}.
This is expected mainly in systems with a neutron star component, where natal kicks at birth can impart significant orbital eccentricity
\citep[e.g.][and references therein]{2025arXiv250901430P, 2025arXiv250508857V, 2025A&A...701L...3V}.

Disentangling Galactic gravitational-wave sources is an essential first step in interpreting LISA data and is a necessary prerequisite for addressing binary-type–specific LISA science goals \citep{Redbook}. In this work, we investigate whether machine learning techniques can aid in overcoming the expected difficulties and enable accurate classification of Galactic stellar remnant binaries in LISA data. Specifically, we explore the use of machine learning algorithms for classifying multiple binary classes detectable by LISA. In addition, by focusing on two largely overlapping classes (WDWD and NSWD), we assess the strengths of these algorithms and compare their performance with that of a classical statistical approach based on the kernel density estimation
(KDE). Finally, we discuss potential other applications of such classification that can advance LISA science.

The remainder of this paper is organised as follows. In \cref{sec: methods}, we describe the gravitational-wave parameters, the synthesis of the catalogues, and how they were used for training and testing our machine learning algorithms. We also outline the types of classifiers employed, the methods applied, and the metrics used to evaluate their performance. The main results are presented in \cref{sec:results}, starting with those of the multi-class classifier, followed by an analysis of the performance of other classifiers in the binary classification study. In \cref{sec:other app}, we discuss the potential of extending this classification approach to other use cases in the LISA context. Finally, we conclude our work in \cref{sec:conclusion}.

\section{Methods}\label{sec: methods}

In this section, we describe the definitions of the gravitational-wave features and astrophysical catalogues used for training and testing our classifiers. We then outline the classifiers employed in our study, along with the optimisation and probability calibration methods applied. Finally, we define the evaluation metrics used to assess the performance of our classifiers.

\subsection{Binary's detectability with LISA} \label{sec: GW signal}

Consider a binary system with masses $m_1$ and $m_2$ and semi-major axis $a$.
Kepler’s law relates the orbital period to the total mass $M=m_1+m_2$; expressed in terms of the gravitational-wave frequency:
\begin{align} \label{eq:KeplerIII}
    f_{\rm GW} &= \frac{1}{\pi} \sqrt{\frac{GM}{a^3}}\\ &= \left(\frac{M}{1\,\mathrm{M}_\odot}\right)^{1/2}\left(\frac{a}{1\,\mathrm{R}_\odot}\right)^{-3/2} \times 0.20\,\mathrm{mHz} .
\end{align}
Gravitational-wave radiation causes the orbit to shrink, producing a gradual chirp of the frequency:
\begin{align} \label{eq:chirp}
    \dot{f}_{\rm GW} &= \frac{96}{5\pi c^5}(G\mathcal{M}_c)^{5/3}(\pi f_{\rm GW})^{11/3}\\ &= \left(\frac{\mathcal{M}_c}{1\,\mathrm{M}_\odot}\right)^{5/3}\left(\frac{f_{\rm GW}}{1\,\mathrm{mHz}}\right)^{11/3} \times 0.18\,\mathrm{nHz}\,\mathrm{year}^{-1},
\end{align}
where $\mathcal{M}_c$ is the chirp mass. A binary’s detectability with LISA is primarily determined by its GW strain amplitude:
\begin{align} \label{eq:GWamp}
    \mathcal{A}_0 &=  \frac{2(G\mathcal{M}_c)^{5/3}(\pi f_\mathrm{GW})^{2/3}}{D c^4} \\
    &= \left(\frac{\mathcal{M}_c}{1\,\mathrm{M}_\odot}\right)^{5/3} \left(\frac{f_\mathrm{GW}}{1\,\mathrm{mHz}}\right)^{2/3} \left(\frac{10\,\mathrm{kpc}}{D}\right) \times 5.94\times 10^{-23} .
\end{align}
Since gravitational-wave instruments respond directly to the strain amplitude—rather than to flux, as in electromagnetic astronomy—the corresponding signal-to-noise ratio (SNR) over an observation time $T_{\rm obs}$ is 
\begin{equation} 
\langle \rho^2 \rangle
\propto \frac{\mathcal{A}_0^2 T_{\rm obs}}
{S_n(f_{\rm GW})}, \label{eq:snr}
\end{equation}
where $S_n(f)$ is LISA’s noise spectral density defined in the \citet{LISASciRD}.

Current LISA global fit pipelines model Galactic binaries using circular, quasi-monochromatic waveforms characterised by eight parameters:
\begin{equation}
    \{ f_{\rm GW}, \dot{f}_{\rm GW}, \mathcal{A}_0, \lambda, \beta, \iota, \psi, \phi_0 \},
\end{equation}
where, in addition to the quantities defined above, $(\lambda, \beta)$ are the sky coordinates, $\iota$ is the inclination, $\psi$ the gravitational-wave polarisation angle, and $\phi_0$ the initial orbital phase \cite[see][]{Littenberg2023, Strub2024, Katz2025, Deng2025}. These parameters are direct LISA observables and therefore form the core feature set used in our machine learning analysis.
We also include the SNR ($\rho$) as an additional feature, as it will likely be reported in future LISA source catalogues and may carry useful information for training.
Recent studies have begun to incorporate eccentricity into waveform models and parameter estimation of Galactic LISA sources \citep[e.g.][]{Katz2022, Moore2024, Middleton2025}. Motivated by these studies and by the astrophysical importance of eccentricity, we include eccentricity $e$ as an additional feature in our analysis.

For completeness, we note that, if the binary is eccentric, gravitational-wave radiation is emitted across multiple harmonics of the orbital frequency \citep{PetersMathews1963}.
Each harmonic contributes power at
\begin{equation}
f_n = n f_{\rm orb} = \frac{n f_{\rm GW}}{2},
\end{equation}
with strain amplitude
\begin{equation}
\mathcal{A}_n =
\mathcal{A}_0\left(\frac{2}{n}\right)^{5/3} \sqrt{g(n,e)},
\end{equation}
where $g(n,e)$ is defined in \citet{{PetersMathews1963}}.
In this case, the total SNR is obtained by summing the SNRs of the detectable harmonics in quadrature.

\subsection{Main catalogue of LISA-detectable Galactic binaries for training and testing}\label{sec: main catalogue}

To train our machine learning classifiers, we construct a representative catalogue of Galactic sources detectable by LISA.
Our catalogue combines white dwarf binaries (the `low-mass' population of WDWD and NSWD) with neutron star and black hole binaries (the `high-mass' population of NSNS, BHNS, BHBH), drawn from two separate published population studies. As a result, the populations are not generated within a single self-consistent framework; however, for the purpose of our exploratory work—focusing solely on disentangling binary types—this approach is sufficient. We ensure that the relative numbers of each binary class reflect astrophysically motivated, fiducial expectations for LISA detections. 

The low-mass population, comprising the WDWD and NSWD binaries, is taken from our earlier work \citep{Korol2024}. In that study, the populations were generated using the \texttt{SeBa} binary population synthesis code \citep{SeBa, Nelemans2001,Toonen2012, Toonen2018}, which is widely used for modelling low-mass binary evolution and is calibrated against observed samples \citep{Nelemans2000,Nelemans2005,vanderSluys2006,too17}. The population comprises a suite of nine variations, each based on different assumptions about binary interactions (specifically the treatment of common-envelope evolution) and the distribution of natal kicks imparted to neutron stars at birth. Since these binary classes are expected to constitute $\sim$99.9\% of the LISA-detectable Galactic population, they are the most critical to model self-consistently, particularly to separate individually detectable binaries from those that contribute to the unresolved Galactic foreground.
For this purpose, all populations variations were processed through an iterative pipeline described in \citet{Karnesis2021}, which approximates the LISA global fit analysis.

Based on the SNR threshold, the pipeline iteratively separates individually resolved (high-$\rho$) sources from the unresolved (low-$\rho$) Galactic population, allowing us to extract a subset of LISA-detectable binaries for each catalogue variation. We note that the pipeline employs purely circular waveforms. Given that a significant fraction ($\sim 45\%$) of NSWD binaries in our mock catalogues retain non-zero eccentricities, each eccentric NSWD catalogue entry is decomposed into a collection of harmonics (see Sect.~\ref{sec: GW signal}), with the number of harmonics chosen to capture $\sim 99\%$ of the emitted energy. However, only up to several harmonics of a given binary are detectable across all mock catalogue variations (see Figure~4 of \citealt{Korol2024}). We retain all detectable harmonics for training as they can be detected as separate individual sources, even though they remain physically correlated. The resulting low-mass component in the main catalogue of LISA detectable binaries contains a total of 50,255 binaries, of which 48,311 are WDWD systems and 1,944 are NSWD systems of which 1,410 of them are unique (i.e. not representing harmonics of the same eccentric NSWD system).

\begin{table}
\caption{Gravitational-wave features used for machine learning classification.}
\label{tab:prior_of_catC} \label{tab:features}
\centering
\begin{tabular}{llll}
\hline \hline
Parameter   & Symbol             & Units & Origin \\
\hline
GW frequency              & $f_{\rm GW}$       & Hz       & K24/W22 \\
Frequency derivative      & $\dot{f}_{\rm GW}$ & Hz/s     & K24/W22 \\
Amplitude                 & $\mathcal{A}_0$      & ---      & K24/W22 \\
Eccentricity              & $e$                & ---      & K24/W22 \\
Signal-to-noise ratio     & $\rho$             & ---      & K24/W22 \\
Ecliptic longitude        & $\lambda$          & rad      & K24/W22 \\
Ecliptic latitude         & $\beta$            & rad      & K24/W22 \\
Inclination angle         & $\iota$            & rad      & $\cos\iota \sim \mathcal{U}(-1, 1)$ \\
Polarisation angle        & $\psi$             & rad      & $\mathcal{U}(0, 2\pi)$ \\
Initial phase             & $\phi_0$           & rad      & $\mathcal{U}(0, 2\pi)$ \\
\hline
\end{tabular}
\tablefoot{Values for $f_{\rm GW}$, $\dot{f}_{\rm GW}$, $\mathcal{A}$, $e$, $\lambda$, and $\beta$ for the WDWD and NSWD binary classes are taken directly from the astrophysical catalogues of \citet[][K24]{Korol2024}, while the corresponding values for the NSNS, BHNS, and BHBH classes are taken from \citet[][W22]{Wagg2022}.
The remaining angular parameters are drawn assuming isotropy. Here, $\mathcal{U}(a,b)$ denotes a uniform distribution between $a$ and $b$.}
\end{table}

\begin{figure}
\centering
\includegraphics[width=\hsize]{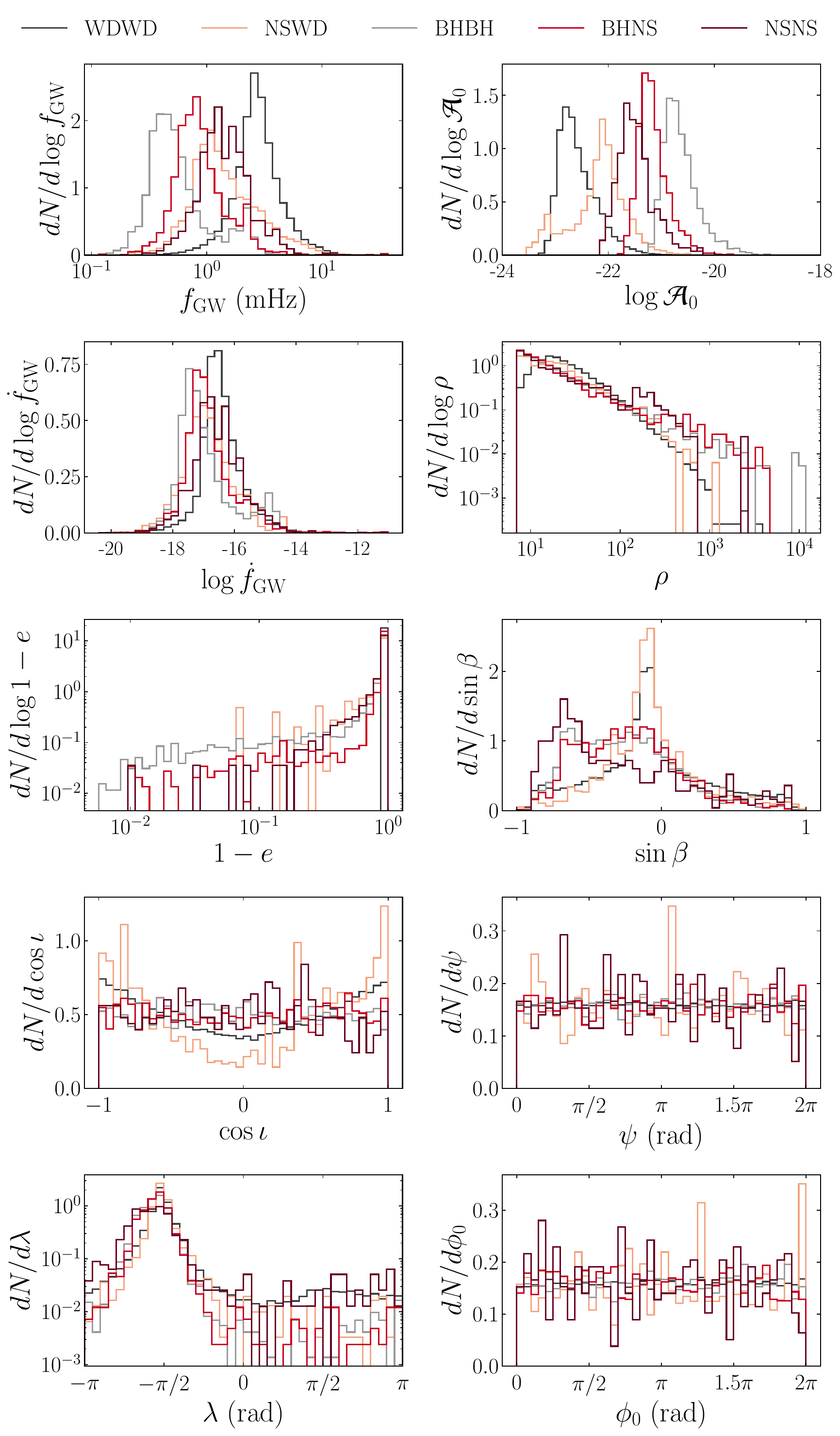}
\caption{Distributions of the ten features in the data used for the machine learning classifiers, coloured by true class. Black: WDWD binaries; orange: NSWD binaries; grey: BHBH binaries; red: BHNS binaries; and maroon: NSNS binaries. Each distribution is normalised independently.}
\label{fig:Training-distribution}
\end{figure}

The high-mass population consists of NSNS, BHNS, and BHBH binaries drawn from the fiducial model of \citet{Wagg2022}, obtained using the \texttt{COMPAS} population synthesis code \citep{Broekgaarden2021, Broekgaarden2022, COMPAS}. For each of these binary classes, we select a weighted random subset of systems according to the simulation weights.
In \citet{Wagg2022}, the fiducial population model predicts that, over a four-year (ten-year) LISA mission, at least 37 (70) BHBHs, 18 (38) BHNSs, and 5 (8) NSNSs would be distinguishable above the WDWD foreground. To preserve these relative proportions while ensuring sufficient training statistics, we inflate the absolute counts and sample (4,630; 2,630; 500) binaries for BHBH, BHNS, and NSNS respectively. This yields a dataset large enough for machine learning applications. 

To preserve an astrophysically motivated, fiducial test set, we drew and reserved a representative subset of the high-mass binaries: 93 BHBH, 53 BHNS, and 10 NSNS systems (roughly matching the expected LISA detection counts from \citealt{Wagg2022}). 

For the low-mass binaries, we reserved 20\% for testing, resulting in 9,662 WDWD and 389 NSWD systems. 
We then combined the low- and high-mass populations. All remaining data (i.e., not reserved for testing) was split 80/20 into training and validation sets.

We summarise gravitational-wave parameters (features) used for the training in Table~\ref{tab:features} and show the distribution of the features in the training data in \cref{fig:Training-distribution}. Finally, we note that the input features were standardised through z-score normalisation based on the training data, and the same transformation was applied to the validation and test sets to ensure consistency.

\subsubsection{Other auxiliary catalogues}\label{sec:other catalogues}

In addition to our main catalogue, we consider a single random realisation of the low-mass binary population, based on one of the astrophysical model variations included in the main catalogue. By construction, this realisation preserves the total number and relative proportions of each class expected from a realistic four-year LISA observation: 20,791 binaries in total, comprising 20,608 WDWD systems and 183 NSWD systems. Because this catalogue is used exclusively for performance evaluation, we will refer to it as the low-mass test catalogue.

We also consider an independent catalogue of low-mass binaries from \citet{Korol2022Obs}, which was also not used during training. Importantly, it contains only WDWD systems, with a total of 59,250 binaries. We use this catalogue to assess the classifier’s ability to correctly identify a population consisting solely of WDWD systems, and demonstrate robustness when applied to data derived from an unseen astrophysical model. This catalogue represents a stringent test of whether the classifier can generalise beyond the assumptions built into the training data.

As for the main catalogue, the input features in both auxiliary catalogues were also normalised using the transformation derived from the training data, and the same transformation is applied to the validation and test sets, to the low-mass test catalogue, and to the independent catalogue.

\subsection{Machine learning classification methods}\label{sec: ML class}

In this study, we adopt an explorative approach to identify the most suitable classifier for distinguishing among different types of compact binaries. To this end, we evaluate seven machine learning techniques, ranging from simple baseline models to more complex probabilistic and ensemble methods, and one simple statistical approach. 

The classifiers used in our analysis are:
\begin{itemize}
    \item K-Nearest Neighbour \citep[KNN;][]{KNN1951, KNN1967}: One of the simplest machine learning classification techniques using a distance-based method, that provides a baseline for how well local neighbourhood information alone can separate classes.
    \item Support Vector Machine \citep[SVM;][]{svm1995}: Effective at finding separating hyperplanes in high-dimensional feature spaces, making it suitable for testing linear class boundaries.
    \item Random Forest \citep[RF;][]{RF2001}: An ensemble of decision trees that aims to reduce overfitting through bagging and random feature selection.
    \item Deep Neural Network (NN): A custom made flexible, high-capacity model that can capture complex, non-linear relationships between input and output parameters (for more details see Appendix~\ref{sec:binaryclassifier}).
    \item Natural Gradient Boosting \citep[NGBoost;][]{duan2020ngboostnaturalgradientboosting}: A method for probabilistic prediction that produces calibrated uncertainty estimates.
    \item Gaussian Mixture Model (GMM): 
    Each class is modelled as a mixture of multivariate Gaussian components in the feature space by maximising the likelihood of the training data for that class. 
    We follow the implementation of  \cite{Bailer_Jones_2019} but incorporate our class priors to address class imbalance. 
    \item XGBoost \citep{xgboost2016}: 
    A highly optimized gradient-boosted decision tree algorithm known for its strong performance and efficiency across structured datasets, making it a natural candidate for our tabular features. 

\end{itemize}
Details of the classifier implementations are provided in \cref{app: details of implementation}. All classifiers were optimised using Bayesian optimisation (see \cref{sec:bayes_opt}), and their probabilistic outputs were calibrated using a Bayes’ rule–based method adapted from \cite[][for details see \cref{sec:bayesrule}]{Berbel_2024}. 

We compare the classifiers listed above to
a simple statistical approach to classification based on KDE combined with Bayes’ theorem. Let $\mathcal{D}$ denote the training dataset consisting of ten gravitational-wave features $\vec{x}$ listed in Table~\ref{tab:features}. We used a Gaussian kernel, with the bandwidth selected via grid search. 
Specifically, we explored a logarithmically spaced range of values given by \texttt{np.logspace(-1, 1, 20)}, which yielded an optimal bandwidth of $h = 0.20691$. 
Using the KDE, we approximated the class-conditional densities.
$\hat p(\vec{x} \mid y=0)$ and $\hat p(\vec{x} \mid y=1)$ where they represent the likelihood of class 0 and class 1 given a set of features $\vec{x}$ respectively.
Together with the empirical class priors $\pi_0$ and $\pi_1$ (in our case these are simply class ratio estimated from our catalogues), estimated from $\mathcal{D}$. Bayes’ theorem then yields the posterior probability for class 1:
\begin{equation}
P(y=1 \mid \vec{x}) = \frac{ \hat p(\vec{x} \mid y=1) \pi_1}{ \hat p(\vec{x} \mid y=1)\pi_1  + \hat p(\vec{x} \mid y=0) \pi_0}.
\end{equation}
The posterior for class 0 follows analogously.

\subsubsection{Evaluation metrics}\label{sec:evaluation metric}

When presenting our results, we use confusion matrices to illustrate the performance of each classifier. 
Here, each row represents the true class and each column the predicted class; correct classifications appear on the diagonal, while off-diagonal elements represent misclassifications.

When examining the low-mass population by itself,  where there are only 2 possible classes (WDWD and NSWD), we also evaluate the classifiers' performance using the specificity, sensitivity, Matthews correlation coefficient (MCC), and the area under the receiver operating characteristic (AUROC) curve.
In this binary classification task, we define the positive class to be NSWD and the negative class to be WDWD.

The specificity (also known as the true negative rate, TNR) is defined as
\begin{equation}
    \text{Specificity} = \frac{\text{TN}}{\text{TN} + \text{FP}}.
\end{equation}
This metric is an indication of how well the classifier predicts the negative class. Its values range from 0 to 1, with values closer to 1 indicating stronger performance.

Sensitivity (or also known as the true positive rate, TPR, or recall) is defined as
\begin{equation}
    \text{Sensitivity} = \frac{\text{TP}}{\text{TP} + \text{FN}},
\end{equation}
and quantifies how well the classifier recovers the positive class. As with specificity, values closer to 1 correspond to better performance.

The MCC is defined as:
\begin{equation}
    \text{MCC} =  \frac{\text{TP} \times \text{TN} - \text{FP} \times \text{FN}}{\sqrt{(\text{TP} + \text{FP})(\text{TP} + \text{FN})(\text{TN} + \text{FP})(\text{TN} + \text{FN})}},
\end{equation}
and provides a balanced measure of classification performance. We highlight that the MCC incorporates all four entries of a binary confusion matrix, making it particularly suitable for imbalanced datasets—such as ours, where in the low-mass population, WDWDs are far more numerous than NSWDs. MCC values range from $-1$ (complete disagreement) through $0$ (random performance) to $1$ (perfect classification).

Finally, the AUROC is computed as the area under the curve obtained by plotting sensitivity against specificity. It can be interpreted as the probability that the classifier assigns a higher score to a randomly chosen NSWD binary than to a randomly chosen WDWD. An AUROC value of 0.5 corresponds to random guessing, while a value of 1 indicates perfect discrimination.


\section{Results}\label{sec:results}

In this section, we begin with a multi-class analysis to characterise the overall separability of the various binary systems that are detectable by LISA in the Milky Way. 
We then focus on binary classification between the most numerous and also the most challenging classes, (namely the low-mass population of WDWD and NSWD), with the aim of quantifying their separability and evaluating the performance of different machine learning classifiers. After identifying the best-performing classifiers, we proceed to optimise their performance using Bayesian optimisation (see \cref{sec:bayes_opt}) and calibrate their outputs with a Bayes’ rule-inspired probability calibration method (see \cref{sec:bayesrule}). Finally, we analyse which features drive the classification, and we assess the classifier’s performance on both a realistic test catalogue and an independent catalogue to evaluate its robustness under LISA-like observational conditions.

\subsection{Performance on multi-class test set}

\begin{figure}
\centering
\includegraphics[width=\hsize]{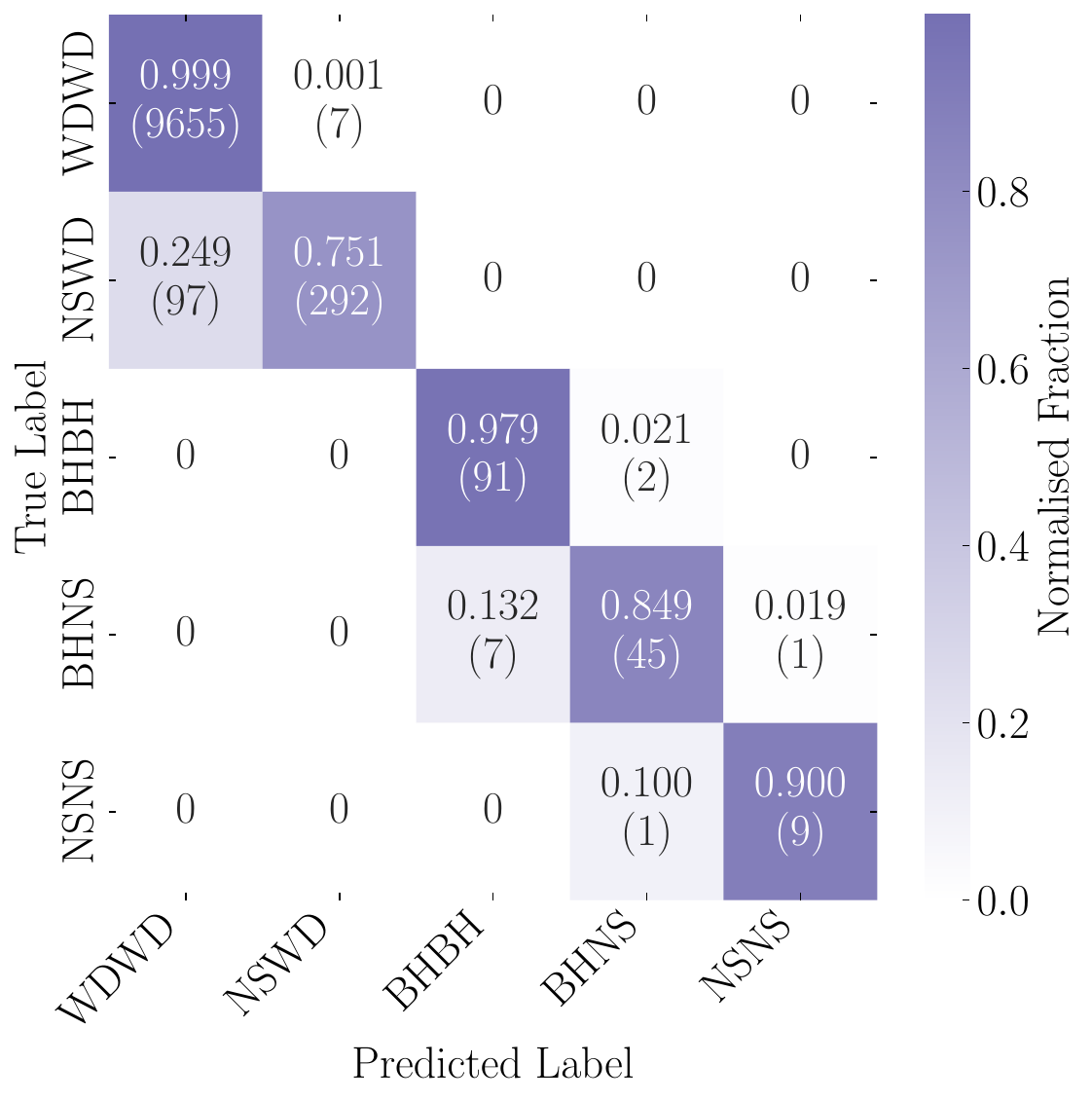}
  \caption{Confusion matrix of the multi-class XGBoost classifier evaluated on the main catalogue’s test set. Each entry is row-normalised and colour-coded by value, with bracketed numbers indicating the absolute counts. The classifier demonstrates strong performance in separating the high-mass binaries (BHBH, BHNS, NSNS), while classification of the low-mass binaries (WDWD, NSWD) remains more challenging. The distinction between WDWD and NSWD is the most difficult, with approximately $25\%$ of NSWD class incorrectly predicted as WDWD.
          }
     \label{fig:multiclass-confusion-matrix}
\end{figure}

For the multi-class analysis, we use XGBoost, which showed the strongest performance in preliminary tests. We train it on the main catalogue described in \cref{sec: main catalogue}, which includes five binary classes (WDWD, NSWD, BHBH, BHNS, and NSNS), using the training set shown in Fig.~\ref{fig:Training-distribution}.
We then optimise the model using Bayesian optimisation, with the resulting hyper-parameters listed in 
\cref{app: hyper}.
Finally, we evaluate the trained classifier on the main catalogue’s test set.

\cref{fig:multiclass-confusion-matrix} shows the performance of the XGBoost classifier on the main catalogue’s test set. The confusion matrix indicates that the classifier achieves high recall for the high-mass binaries (BHBH, BHNS, and NSNS), with correctly classified fractions exceeding 85\% for all three classes. The WDWD systems are also identified with a recall of 99.9\%, demonstrating the model’s ability to capture the dominant population effectively. However, separating NSWD systems from WDWD binaries proves substantially more challenging: about 25\% of true NSWDs are misclassified as WDWDs. This difficulty arises from the close similarity of the gravitational-wave features of WDWD and NSWD binaries within the LISA band. Both classes occupy overlapping regions in frequency, amplitude, and chirp rate, making their signals nearly indistinguishable without an additional measurable handle such as eccentricity (see Fig.~\ref{fig:Training-distribution}).

In addition, the extreme class imbalance naturally biases the classifier toward predicting the WDWD class, further increasing NSWD$\to$WDWD misclassifications. 
Although class-balancing techniques could mitigate this, they may come at the cost of reduced precision for the dominant WDWD population.

\subsection{Performance of classifiers on low-mass population}\label{sec:perf_binary}

\begin{figure}
\centering
\includegraphics[width=\hsize]{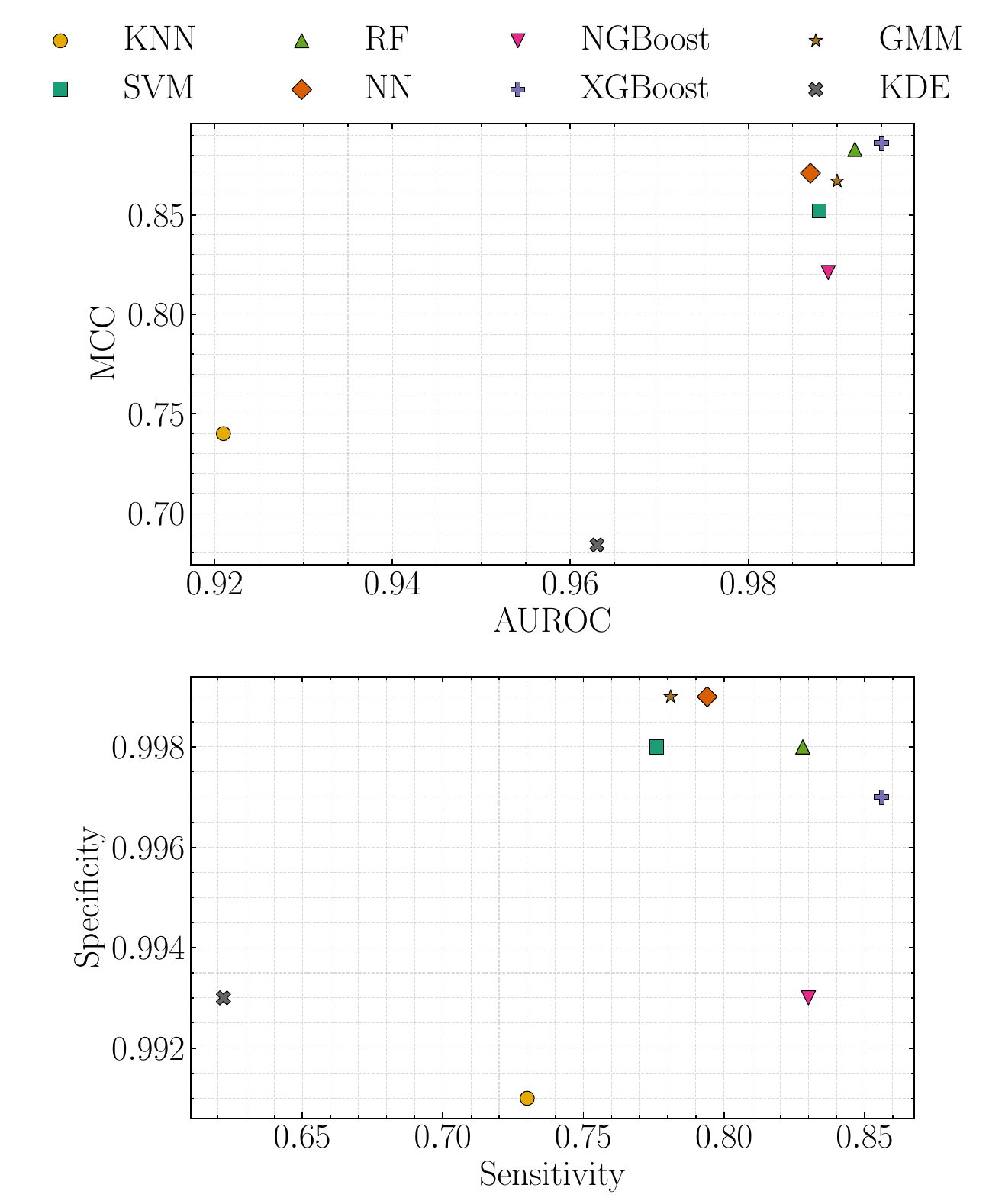}
  \caption{Performance comparison of machine-learning classifiers evaluated on the test set including NSWD and WDWD binaries only. 
  In both plots, the top-right corner represents the ideal performance. 
          }
     \label{Fig:model_comparison}
\end{figure}

The multi-class results highlight that separating the most numerous classes, namely WDWD from NSWD systems, is the dominant challenge.
Focusing on this low-mass population, we trained the 8 classifiers of \cref{sec: ML class} on WDWD and NSWD binaries only and evaluated them on the matching test set. We then calibrated the probability outputs (see \cref{sec:bayesrule}).

After this calibration, each model provides a continuous probability score $p$ of a system being NSWD rather than WDWD. However, since the two classes show considerable feature-space overlap, the optimal decision threshold $s$ is not guaranteed to be $0.5$. To maximize classification performance, we optimized $s$ for each classifier to maximize MCC on the test data. The resulting decision rule is

\begin{equation}\label{eqn:prediction}
\hat{y} =
\begin{cases}
 0 \;\;(\text{WDWD}), & \text{if } p \leq s, \\
1 \;\;(\text{NSWD}), & \text{if } p > s,
\end{cases}
\end{equation}
The determined threshold values, listed in ~\cref{tab:score threshold}, varied across methods. Notably, some classifiers (e.g., KNN and NGBoost) required substantial shifts from $0.5$. This reflects factors such as class imbalance, feature-space overlap, or the classifier's intrinsic probability scaling,
highlighting the importance of threshold optimization in our specific problem.

The performance of all 8 classifiers on the low-mass population can be seen in 
\cref{Fig:model_comparison}.

The figure presents four key metrics introduced in Section~\ref{sec:evaluation metric}: AUROC, MCC, sensitivity, and specificity, with the ideal performance located in the upper-right region of each panel. 
Across all models, the specificity is very high (typically $\gtrsim 0.99$), meaning that WDWD systems are reliably identified. The variation arises primarily in the sensitivity, where KDE and KNN perform noticeably worse.
All classifiers achieve AUROC values above 0.92 and positive MCC scores greater than 0.5, confirming that every method performs well above random and retains real discriminatory power despite the strong class imbalance.

Among the tested algorithms, XGBoost, RF, and GMM cluster in the upper-right region of both panels, indicating superior and well-balanced performance compared to the other methods in identifying both classes. Our custom NN performs competitively, showing the effectiveness of classifiers that are capable of capturing non-linear relationships among gravitational-wave features. In contrast, KNN and KDE lie significantly farther from the ideal region, confirming that simple distance-based and non-parametric density estimators struggle with the overlapping, highly imbalanced feature space of compact binaries. The strong performance of decision-tree–based methods (RF, XGBoost) suggests that the gravitational-wave feature space is particularly well suited to classifiers that exploit hierarchical, non-linear decision boundaries.
Finally, we note that Bayesian optimisation generally improved the classifiers’ performance, pushing them closer to the upper-right region, although it did not significantly alter their relative ranking.

\subsection{Performance of binary classifier - XGBoost}

\begin{figure*}[h!]
\resizebox{\hsize}{!}
        {\includegraphics[width=\hsize]{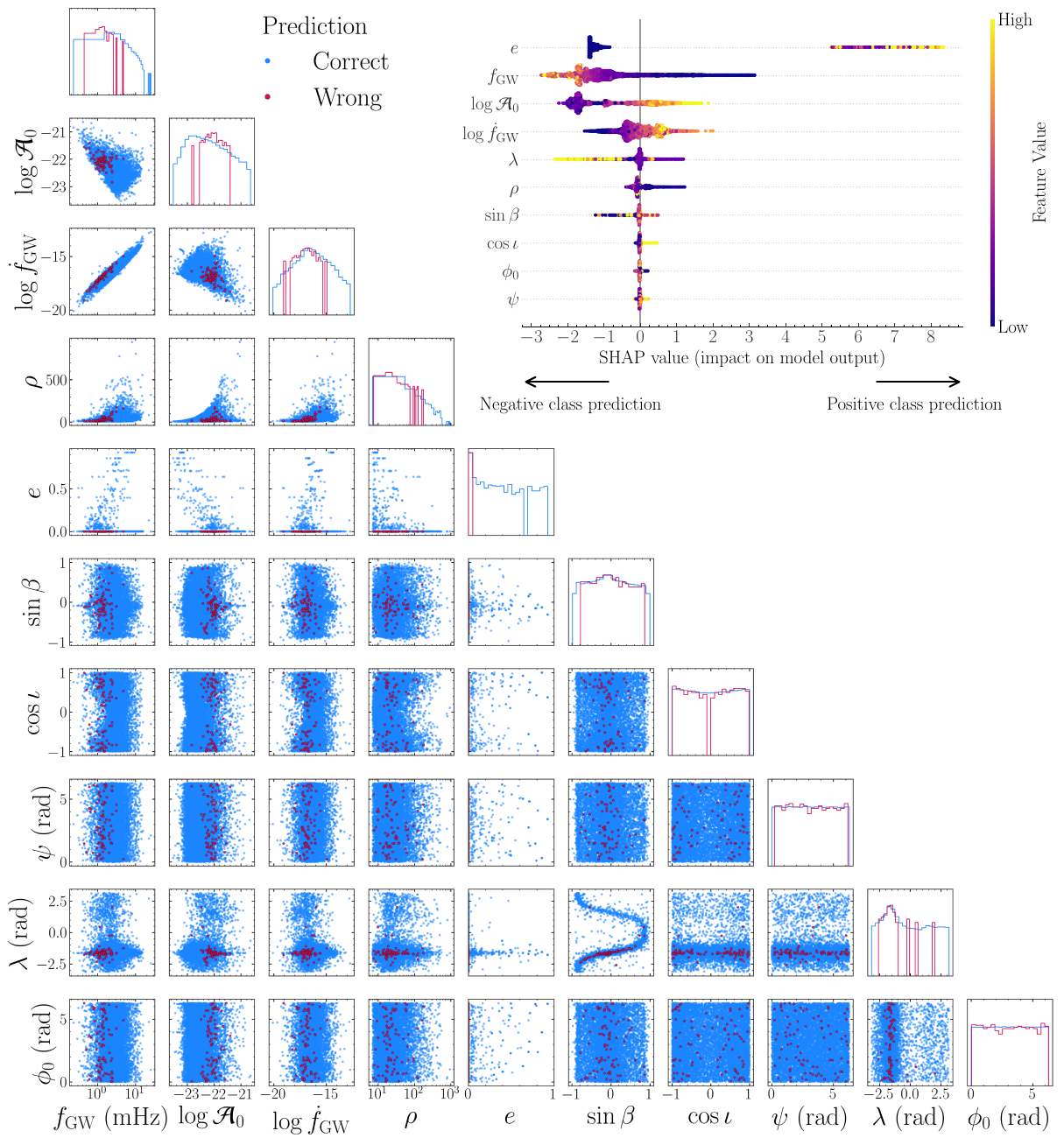}}
    \caption{Feature distributions for correctly (blue) and incorrectly (red) predicted low-mass binary systems (WDWD and NSWD) by XGBoost binary classifier evaluated on the main catalogue's test set. 
    The inset in the upper-right corner shows a SHAP summary plot illustrating the impact of the ten input features on the classifier's output for each system in the low-mass component test set. The features are ranked on the y-axis in descending order of average absolute importance. The x-axis shows the SHAP value, indicating the feature's contribution to the output, where a positive value pushes the prediction towards the positive class (NSWD) and a negative value pushes it towards the other (WDWD). A SHAP value of 0, marked by the vertical grey line, represents the baseline and indicates the feature had no impact on that specific prediction. Each point corresponds to an individual system from the test set, coloured by its normalised feature value, from low (dark blue) to high (yellow).}
     \label{fig:corner_xgboost_test}
\end{figure*}

Since XGBoost demonstrated the best overall performance, we now examine its results in more detail to understand the factors behind its effectiveness and the remaining classification challenges. For context and benchmarking, we also compare its performance to the KDE method, which serves as a simple statistical baseline relying purely on the underlying density distributions without learning complex decision boundaries. The optimised hyper-parameters obtained through Bayesian optimisation for the binary XGBoost classifier are listed in 

\cref{app: hyper}.

The bottom left of \cref{fig:corner_xgboost_test} shows the corner plot of the feature distributions for correctly (blue) and incorrectly (red) predicted compact binary systems with XGBoost. Although the overall feature distributions appear complex, the classifier successfully identifies the majority of binary systems correctly. Among the misclassified systems, clustering is evident within specific ranges of certain features, notably $f_\text{GW}$, $\log \mathcal{A}_0$, $\log \dot{f}_\text{GW}$, and $e$. In contrast, the angular parameters $(\sin\beta, \cos\iota, \lambda, \psi, \phi_0)$ show no clear structure, with the misclassified systems distributed more randomly—except for $\lambda$, where a slight concentration is observed. This behaviour is expected: while $\iota$, $\psi$, and $\phi_0$ are randomly sampled, the sky coordinates follow a Milky Way–like stellar density distribution that peaks toward the Galactic bulge, which can lead to subtle correlations with classification errors, particularly in $\lambda$.

The clustering of misclassified systems in specific regions of the parameter space arises primarily from two factors. First, these regions correspond to high source densities in the astrophysical catalogues, so a larger population naturally yields a higher absolute number of errors.
Second, the intrinsic properties of WDWD and NSWD binaries overlap significantly in these regions, making them inherently difficult to distinguish. 
For reference, in our training set the NSWD-to-WDWD ratio exceeds 0.3 at frequencies below $\lesssim1\,\mathrm{mHz}$, whereas above ${\sim}2$ $\mathrm{mHz}$ the ratio drops to nearly zero.
This implies that the error distribution is at least partly driven by underlying astrophysical structure, although a detailed physical interpretation lies beyond the scope of this work.

\subsubsection{Classifier's interpretability with SHAP}
We can further interpret the performance of a machine learning classifier by using
SHapley Additive exPlanations (SHAP) \citep{lundberg2017unifiedapproachinterpretingmodel}. This is a model interpretation method that utilises the concept of Shapley value in game theory \citep{shapley1953value}. Essentially, SHAP helps to explain how much each input feature contributes to a model's final prediction. 
It assigns a numerical value (SHAP value) for each individual prediction, indicating whether that feature increases or decreases the likelihood of a given outcome.  

This provides an intuitive way to understand the complex decisions made by the model and may also provide clues as to whether or not the features driving its classifications are physically motivated.

The inset in \cref{fig:corner_xgboost_test} on the upper-right plot shows the SHAP summary plot for the binary XGBoost classifier. The features are ranked on the y-axis according to their average absolute SHAP value, indicating their overall importance to the model’s decision. The x-axis represents the SHAP value itself, that is the magnitude and sign of each feature’s contribution to the classification output. Positive values push the prediction towards the NSWD class and negative values towards the WDWD class. Each point corresponds to an individual system in the test set, coloured by the feature’s normalised value.
This analysis reveals that the eccentricity $e$, gravitational-wave frequency $f_\text{GW}$, strain amplitude $\log \mathcal{A}_0$, and frequency derivative $\log \dot{f}_{\text{GW}}$ are the most influential features driving the classifier’s predictions. Specifically, systems with lower $f_\text{GW}$ and higher $\log \mathcal{A}_0$ and $\log \dot{f}_{\text{GW}}$ tend to be classified as NSWDs. For eccentricity $e$, highly eccentric systems are generally classified as NSWDs, whereas systems with zero eccentricity are typically identified as WDWDs. This trend is astrophysically motivated: WDWD binaries are expected to have circular orbits due to tidal circularisation during their common-envelope evolution \citep[e.g.][]{CE,Zahn1977}, while NSWD systems can retain measurable eccentricities as a result of natal kicks imparted during the supernova event that formed the neutron star \citep[e.g.][]{Igoshev2021,2023MNRAS.521.2504O,2025arXiv250508857V}.

Other features, such as SNR ($\rho$) and the angular parameters $(\lambda, \sin\beta, \cos\iota, \psi, \phi_0)$, have comparatively smaller impacts on the classifier’s predictions. This is expected, as the SNR ratio is determined by the strain amplitude, gravitational-wave frequency and frequency derivative (see Eq.~\ref{eq:snr}), meaning it does not provide additional independent information once these features are included in the training data. Similarly, the angular parameters in our catalogue are drawn randomly from their respective distributions (see Table~\ref{tab:features}) for both classes and therefore carry no intrinsic correlation with the binary type. Consequently, these features contribute minimally to the classification process.

\subsubsection{XGBoost vs. KDE-based classification}

\begin{figure}[h!]
    \centering
    \begin{subfigure}[t]{0.7\hsize}
        \centering
        \includegraphics[width=\hsize]{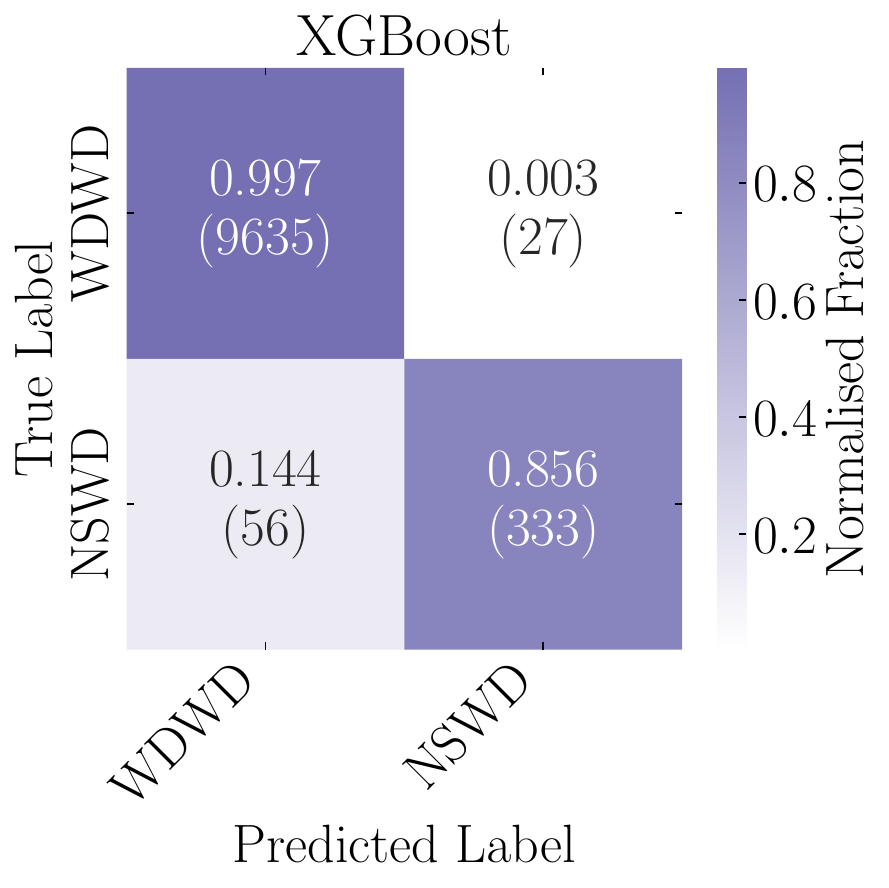}
    \end{subfigure}
    \vspace{1em}
    \begin{subfigure}[t]{0.7\hsize}
        \centering
        \includegraphics[width=\hsize]{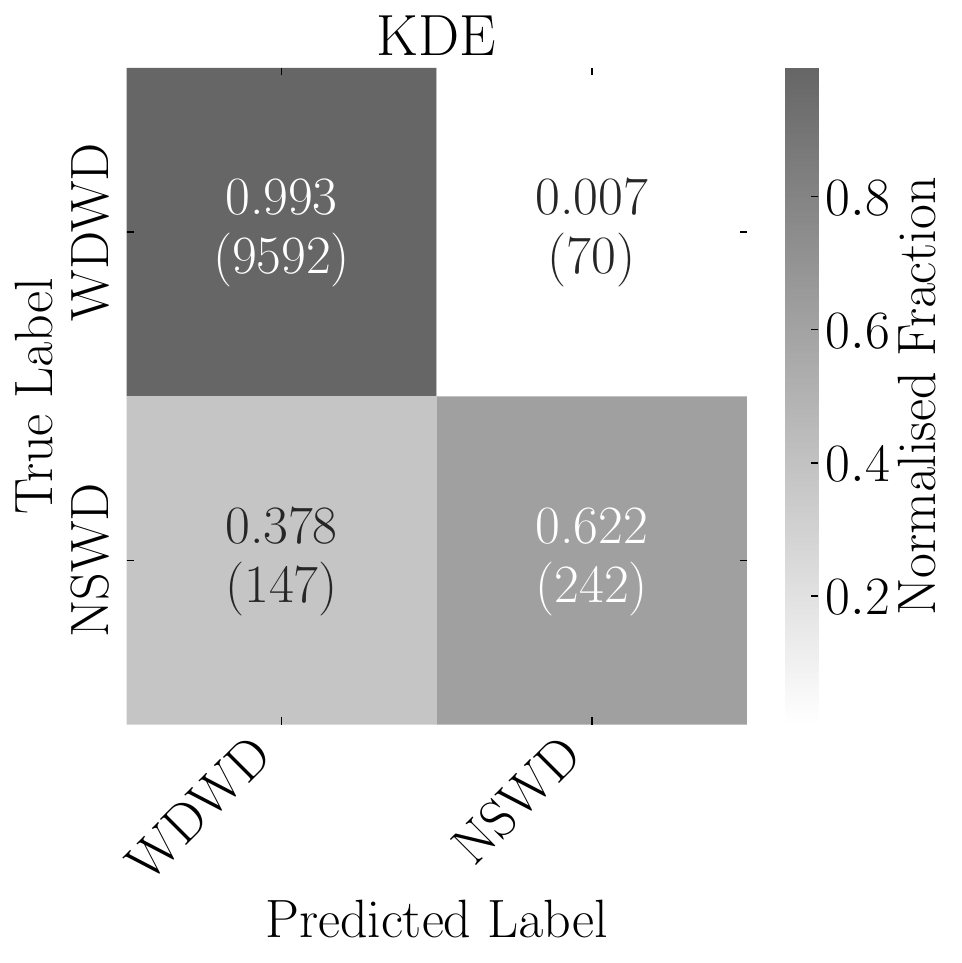}
    \end{subfigure}
    \caption{Confusion matrices evaluated on the low-mass population test set for the XGBoost (purple) and KDE (grey) classifiers. Each entry is row-normalised and colour-coded by value, with bracketed numbers indicating the absolute counts. The results show that XGBoost performs significantly better, correctly predicting 85.6\% of NSWD systems compared to 62.2\% for KDE. For the same population, XGBoost predicts 360 NSWD systems, whereas KDE predicts 312.
    }
    \label{fig:xgboost-kde-comparsion-cm}
\end{figure}

\begin{figure*}[h!]
\resizebox{\hsize}{!}
{\includegraphics[width=1.5\columnwidth]{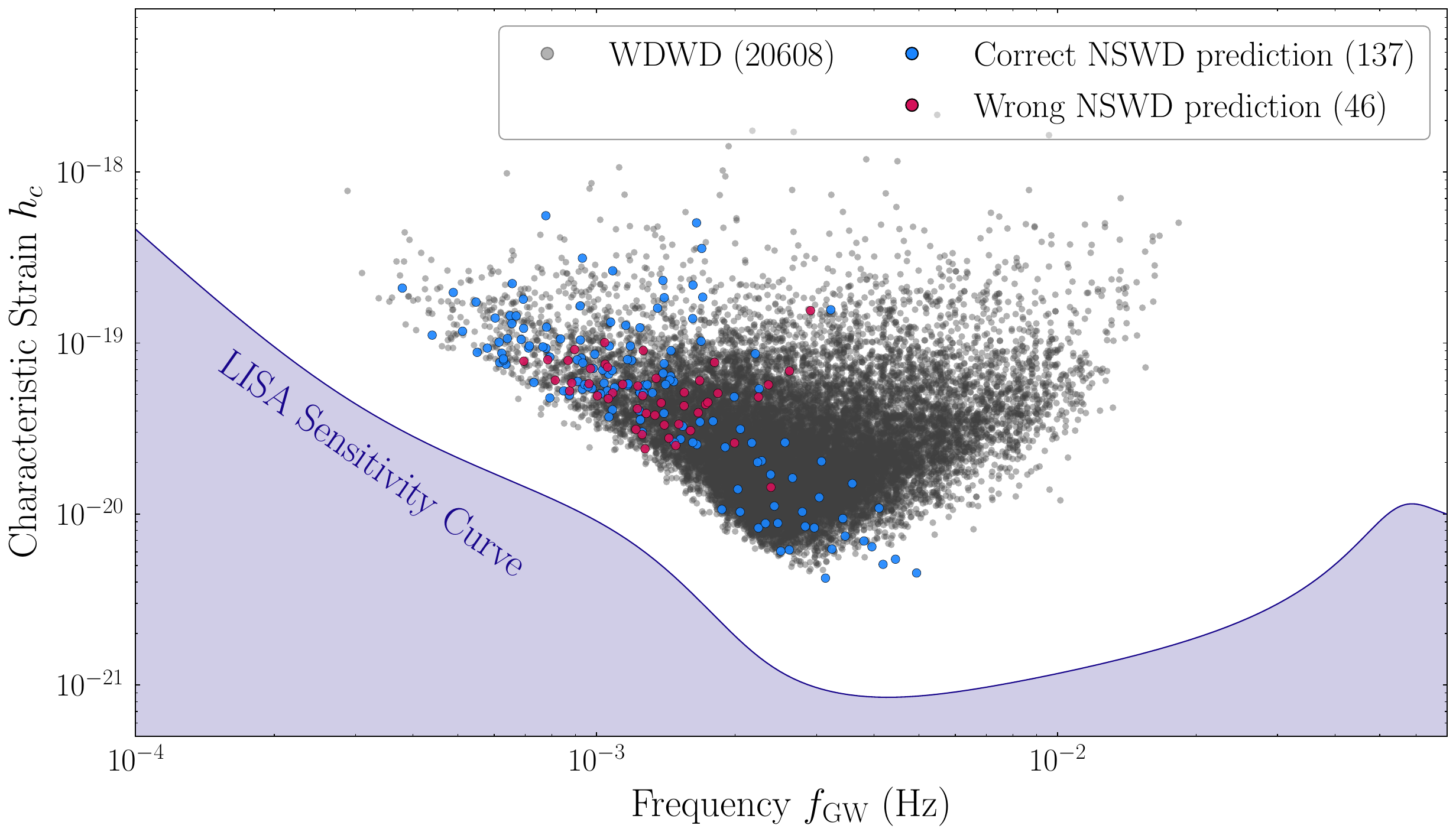}}
\caption{LISA-detectable Galactic WDWD (grey) populations and NSWD (blue and red) populations of a four-year LISA observation period shown in the gravitational-wave frequency–characteristic strain parameter space, with the LISA sensitivity curve plotted and shaded in purple. The two NSWD scatter plots represent the correctly (blue) and incorrectly (red) predicted NSWD systems by the XGBoost classifier evaluated on the test catalogue.}
\label{fig:distributionxtgboostresultmilkyway}
\end{figure*}

\cref{fig:xgboost-kde-comparsion-cm} shows the confusion matrices for XGBoost and KDE on the low-mass population. The KDE method achieves high specificity for WDWD systems (99.3\%) but correctly identifies only 62.2\% of NSWD systems.
 
In contrast, XGBoost substantially improves performance on the minority class, correctly classifying 85.6\% of NSWD systems while maintaining excellent specificity (99.7\%) for WDWD systems. These results confirm that XGBoost achieves a very balanced separation between the two binary populations, effectively mitigating the bias toward the dominant WDWD class while preserving high specificity. 
Furthermore, when compared to the multi-class XGBoost classifier, the binary version shows a clear improvement, with 85.6\% of NSWD systems correctly predicted compared to 75.3\% for the multi-class model. This improvement results from the reduced complexity of the classification task: with only two classes, the model can learn cleaner and more effective decision boundaries, and the optimisation procedure is tailored specifically to distinguishing  NSWD from WDWD binaries.

From this comparison, we show that machine learning approaches trained on mock data outperform classical statistical methods, and that deploying an optimised binary classifier yields better performance than a multi-class classifier. The KDE approach relies on the ratio of class densities estimated from the training sample, however, when the overlap between classes becomes significant, its predictions tend towards randomness. XGBoost, on the other hand, learns hierarchical relationships between features and is able to capture the underlying structure of the data more effectively. As a result, we demonstrate that state-of-the-art machine learning algorithms can outperform simple statistical classifiers in accurately distinguishing compact binary populations.

\subsubsection{Tests with Auxiliary Catalogues}

To evaluate the robustness and astrophysical reliability of our best performing classifier, XGBoost, we applied the trained, optimised, and calibrated XGBoost model to the two auxiliary catalogues introduced in \cref{sec:other catalogues}:
(1) the low-mass test catalogue, which is a random realisation of the same astrophysical WDWD and NSWD population models used in our training set and reflects a realistic four-year LISA observation, and (2) the independent catalogue, which contains only WDWD binaries and is generated using a data-driven model rather than the binary population-synthesis models employed during training.

We find that XGBoost’s performance on the low-mass test catalogue remains robust: out of 183 total NSWD systems, only 46 were misclassified. This demonstrates that the model performs significantly better than random classification, even under extreme class-imbalance conditions—where WDWD systems outnumber NSWD systems by a factor of 112—and is able to reliably recover the minority NSWD population in a realistic astrophysical context.
In \cref{fig:distributionxtgboostresultmilkyway}, we show the classification results, and in the discussion below we focus on the performance for NSWD systems, as the classifier consistently achieves high specificity for WDWD systems. The scatter plot highlights the regions where NSWD systems were correctly (blue) and incorrectly (red) classified. We observe that most misclassified NSWD systems cluster within a specific band of the frequency–characteristic strain plane, between 1 and 2\, mHz. While part of this behaviour can be attributed to the intrinsic overlap of WDWD and NSWD binary parameters in this range, we identify the higher source density in this region as the primary driver of the class confusion. A larger NSWD-to-WDWD ratio in that band naturally leads to more misclassifications for statistical reasons, as previously noted for \cref{fig:corner_xgboost_test}. 
At higher frequencies, the fraction of correctly classified NSWD systems increases. This is because these NSWD binaries are typically eccentric, and each system contributes multiple harmonics to the frequency–strain plane. The presence of eccentricity-related features makes them easier for the classifier to identify, as also indicated by the SHAP analysis in \cref{fig:corner_xgboost_test}.

When applying XGBoost to the independent catalogue, we find that the classifier correctly identified all WDWD systems. This demonstrates that even when the underlying astrophysical model differs from that used during training, the classifier still generalises reliably and does not produce spurious NSWD classifications. Since the true Galactic population of compact binaries remains uncertain, it is entirely possible that LISA will observe systems that fall outside the assumptions encoded in current models. The results here suggest that, provided the training set is sufficiently broad and representative, the classifier remains stable under model variation and is therefore likely to retain reliable performance once LISA observations become available.


\section{Other Applications: identifying eccentric and Galactic Bulge systems}\label{sec:other app}

Beyond distinguishing compact-binary types, we investigate a few additional applications of machine-learning classification to LISA data. Specifically, we explore two such examples: identifying eccentric binaries and identifying millisecond-pulsar–white-dwarf binaries in the Galactic bulge.

\subsection{Identification of eccentric systems}\label{sec: eccaftergolbal}

Our SHAP analysis in \cref{fig:corner_xgboost_test} shows that eccentricity is one of the main features contributing to the classification of binary systems between WDWD and NSWD. However, in reality, this feature may not always be measurable. This could be the case, for example, if only a single harmonic of an eccentric binary is detectable, or because eccentricity is not explicitly included in source-extraction pipelines, as is currently the case.
As already noted, currently published LISA global-fit pipelines use circular waveforms, primarily because astrophysical models predict that the vast majority ($\sim$99\%) of LISA sources will be circular WDWD binaries, and major effort to date has therefore focused on extracting this dominant Galactic population.
Still, even with the global fit setup using circular waveforms for Galactic sources, eccentric systems can be identified in the post-processing stage as clusters of correlated circular signals (see \cref{sec: methods}). Here we explore whether our machine-learning classifier can assist in identifying such eccentric systems under these practical constraints. To this end, we conduct two complementary experiments.

\begin{figure}[h!]
    \centering
    \begin{subfigure}[t]{0.7\hsize}
        \centering
        \includegraphics[width=\hsize]{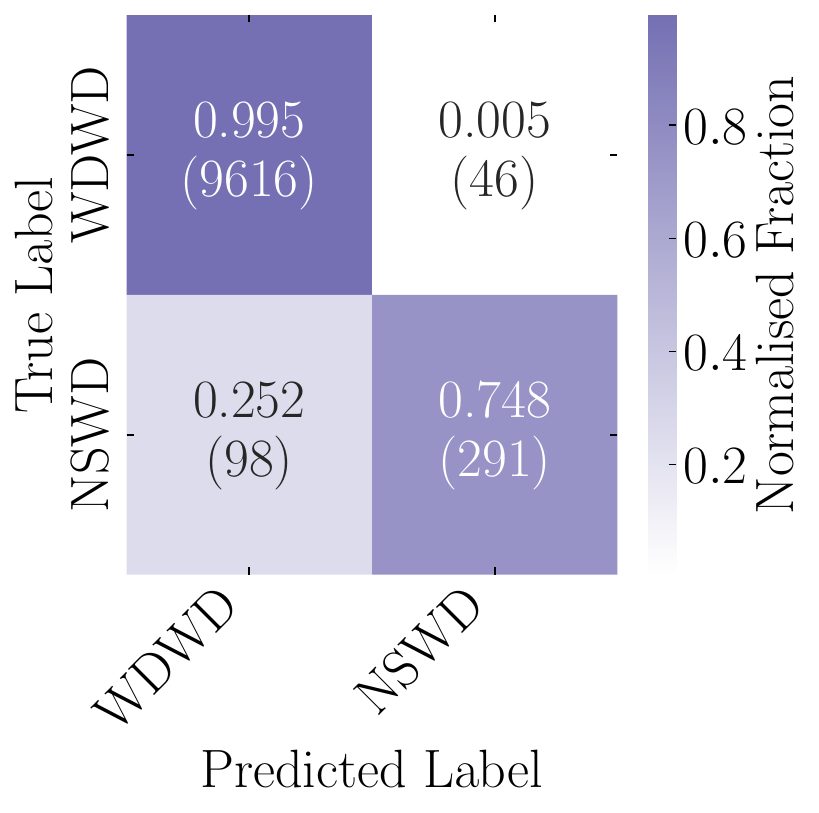}
    \end{subfigure}
    \vspace{1em}
    \begin{subfigure}[t]{0.7\hsize}
        \centering
        \includegraphics[width=\hsize]{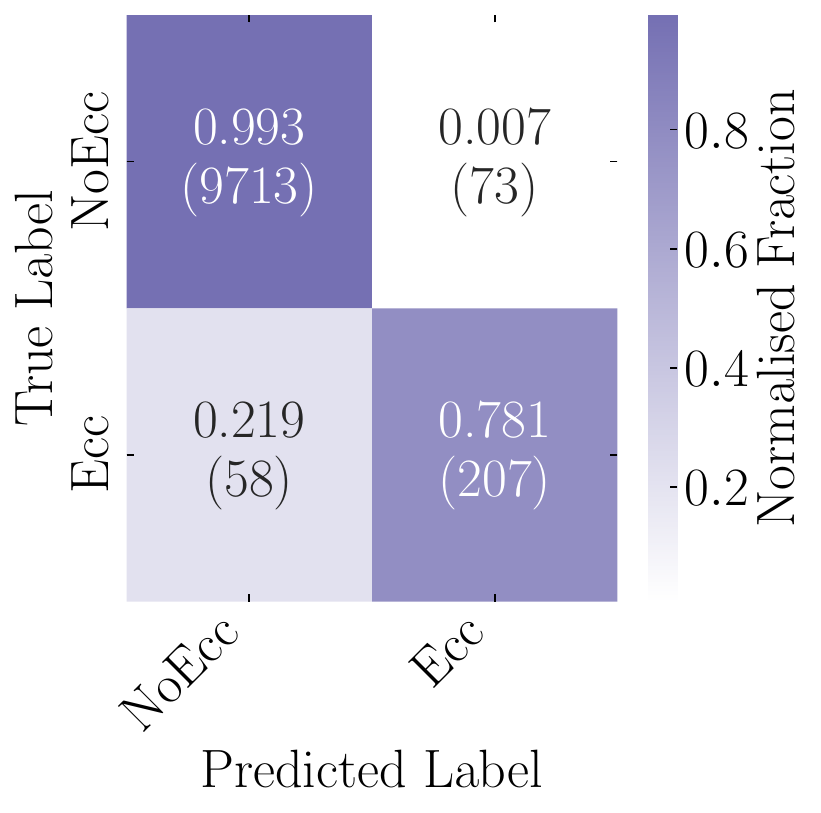}
    \end{subfigure}
    \caption{The confusion matrices for the two experiments described in \cref{sec: eccaftergolbal}. Experiment 1 (top) presents the confusion matrix for the evaluation of the XGBoost classifier trained without the feature eccentricity, applied to classify WDWD and NSWD binaries. Experiment 2 (bottom) shows the confusion matrix for the classifier trained without the eccentricity feature but using labels \texttt{NoEcc} (for binaries with zero eccentricity) and \texttt{Ecc} (for binaries with non-zero eccentricity). Each entry is row-normalised and colour-coded by value, with bracketed numbers indicating the absolute counts.}
    \label{fig:ecc_no_ecc}
\end{figure}

In the first experiment, we train the XGBoost classifier to predict the binary type (WDWD or NSWD) using the same feature set as in \cref{tab:features}, except for leaving out eccentricity. The goal is to assess whether binaries can still be correctly classified based solely on the features available to current global-fit pipelines, which typically assume circular orbits.
The results of this analysis, for the main catalogue’s test set, are presented in the top panel of \cref{fig:ecc_no_ecc}. The classifier correctly identifies 74.8\% of NSWD systems, compared to 85.6\% when eccentricity was included (in \cref{fig:xgboost-kde-comparsion-cm}). 
Although the performance decreases as expected, the results remain encouraging, demonstrating that WDWD and NSWD systems can still be effectively distinguished using the remaining astrophysical parameters alone. 

In the second experiment, instead of predicting the binary type, we train the XGBoost classifier to identify whether a system is eccentric or not, using the labels \texttt{NoEcc} for binaries with zero eccentricity and \texttt{Ecc} for binaries with non-zero eccentricity.
This is illustrated in the bottom panel of \cref{fig:ecc_no_ecc}, where the classifier correctly identifies 207 out of 265 eccentric binaries (78.1\%) and 9713 out of 9786 (99.3\%) non-eccentric binaries.
These results show that even without explicitly including eccentricity as an input feature, the classifier can infer its presence indirectly from correlated quantities such as frequency, amplitude, and frequency derivative. This provides a practical way to flag candidate eccentric sources in global-fit analyses, which could then be subject to more detailed, dedicated follow-up.

\subsection{Identification of millisecond pulsar binaries in the Galactic bulge}

As discussed in \cref{sec:results}, WDWD and NSWD binaries represent the most difficult classes to disentangle, even with state-of-the-art machine-learning classifiers. Within the NSWD population lies an even more challenging subclass: binaries hosting a millisecond pulsar (MSP) and a low-mass white dwarf companion (MSP–WD). These systems form when an old neutron star is spun up through accretion from its companion, producing a rapidly rotating pulsar in a tight, circular orbit with a low-mass WD \citep[for a review, see][]{Lorimer2008LRR,Tauris2011ASPC}. Owing to their low chirp masses and near-zero eccentricities, their gravitational-wave signals are virtually indistinguishable from those of WDWD binaries \citep[][]{korol2025probingmillisecondpulsarorigin}.

Identifying such binaries in LISA data would be particularly valuable, as they represent promising candidates for radio follow-up in future multi-messenger studies. This is especially relevant for the potential population of MSP binaries in the Galactic bulge (within the inner 2–3 kpc), where a GeV $\gamma$-ray excess has been observed \citep{Goodenough2009arXiv,HooperGoodenough2011PhLB,HooperLinden2011PhRvD}. The origin of this excess remains one of the most debated questions in high-energy astrophysics—it could arise from dark matter annihilation or from an unresolved population of MSPs \citep[e.g.][]{2022arXiv220914370H}. Pulsar searches in this region are notoriously challenging due to strong interstellar scattering, high sky temperature, and severe source confusion, which together obscure the underlying MSP population. LISA, however, offers an alternative probe of this population through its gravitational-wave detections \citep[e.g.][]{korol2025probingmillisecondpulsarorigin}.

To explore this possibility, we restricted our catalogue to binaries located within the Galactic bulge and selected only systems with zero eccentricity. 
This assumption is physically motivated, as these systems form when an old neutron star is spun up through accretion, producing a rapidly rotating pulsar in a tight, nearly circular orbit \citep{korol2025probingmillisecondpulsarorigin}.
We then applied our XGBoost classifier to this bulge-only subset. The classifier correctly identifies $8727$ out of $8742$ WDWD systems (99.8\%) and $6$ out of $11$ NSWD systems (54.5\%). Although the identification efficiency for the NSWD systems remains modest, it is still significantly better than random: given the severe class imbalance, a random classifier would select an NSWD with a probability equal to its fraction in the sample, which corresponds to 11 systems out of a total of 8753, or approximately 0.13\%.
Despite these limitations, the results demonstrate that LISA could still identify a fraction of MSP binaries in the Galactic bulge. Even a few such detections would be of considerable interest, as they could confirm the presence of a much larger underlying population and provide valuable targets for coordinated electromagnetic follow-up observations \citep[for details see][]{korol2025probingmillisecondpulsarorigin}. However, there remains potential to improve the classification performance.

\section{Conclusions}\label{sec:conclusion}

In this study, we investigated the feasibility of classifying Galactic stellar-remnant binaries in LISA data using machine-learning techniques. Our aim was to assess whether different Galactic binary types can be distinguished using only gravitational-wave parameters measured by LISA.
Using an XGBoost multi-class classifier, we showed that high-mass binaries (BHBH, BHNS, NSNS) can be readily separated, whereas distinguishing low-mass binaries (WDWD and NSWD) is substantially more challenging and calls for a classifier specifically optimised for this task. Among the methods tested, XGBoost delivered the best performance for this low-mass population, correctly identifying 85.6\% of NSWD systems despite the severe class imbalance. We further demonstrated that the classifier generalises across population synthesis realizations and data-driven models, successfully identifying WDWD systems generated from a different framework than the one used for training. 

We also explored additional astrophysical applications of this approach. First, we showed that the classifier can identify eccentric sources even when eccentricity is not provided explicitly, by learning correlated patterns in other observables. When eccentricity information was removed from the feature set—reflecting the current limitations of LISA global-fit pipelines—the classifier still retained meaningful discriminative power. Second, we applied the classifier to a bulge-restricted catalogue to assess whether LISA could help identify potential MSP-WD binaries in the inner Galaxy. Although classification in this regime remains challenging, the model still identified a non-negligible fraction of NSWD systems, well above random expectation, demonstrating that machine learning may aid in probing Galactic bulge MSP populations.

In summary, our results demonstrate that modern ensemble-based machine-learning algorithms—particularly gradient-boosting methods—offer strong potential for source classification in LISA data as a first step toward the astrophysical interpretation of quasi-monochromatic LISA sources. We have shown that such classifiers can be applied to multiple astrophysical aspects—from distinguishing compact-binary types, to identifying eccentric systems, to pinpointing candidates of rare populations—illustrating how machine learning can facilitate the full scientific potential of LISA.

\begin{acknowledgements}
We thank Silvia Toonen and Tom Wagg for providing the binary population data, and Natalia Korsakova, Stephen Justham, and Christopher Moore for helpful discussions. We also thank the members of the Origins Data Science Lab, in particular Nicole Hartman, Annalena Kofler and Lukas Heinrich, for their feedback on machine learning aspects of this work and valuable suggestions.\\
This research was supported by the Munich Institute for Astro-, Particle and BioPhysics (MIAPbP) which is funded by the Deutsche Forschungsgemeinschaft (DFG, German Research Foundation) under Germany´s Excellence Strategy – EXC-2094 – 390783311.
\end{acknowledgements}

\bibliographystyle{aa} 
\bibliography{biblio} 
%

\begin{appendix} 

\section{Details of implementation}
\label{app: details of implementation}

\subsection{Binary Classifiers}\label{sec:binaryclassifier}

\begin{table*}
\caption{Summary of the hyper-parameter optimisation and probability calibration methods used for each classifier. In general, Bayesian optimisation is adopted as our standard hyper-parameter optimisation approach, and Bayes' rule probability calibration is used as the primary calibration method. 
}\label{tab:classifier-and-methods}
\centering
\begin{tabular}{ccc}
\hline \hline
Classifier & Hyper-parameter optimisation technique & Probability calibration technique  \\
KDE & Grid Search & No calibration\\
KNN & $k=5$ & Bayes’ rule probability calibration\\
SVM & Bayesian Optimisation & Bayes’ rule probability calibration\\
RF & Bayesian Optimisation & Bayes’ rule probability calibration\\
NN & N.A. & No calibration\\
NGBoost & N.A. & No calibration\\
GMM & Bayesian Optimisation & No calibration\\
XGBoost & Bayesian Optimisation & Bayes’ rule probability calibration\\
\hline
\end{tabular}
\end{table*}

 In this section, we elaborate on the methods and parameter choices for the binary classifiers not discussed in the main text. \Cref{tab:classifier-and-methods} lists the optimisation and probability calibration techniques applied to each model. For most classifiers, Bayesian optimisation is used to identify the optimal set of hyper-parameters.  
The exceptions are as follows.
For the KDE method, we used a simple grid search to serve as a baseline.
For the KNN classifier, we selected $k=5$ in order to optimize its $F_1$ score\footnote{$F_1 = \frac{2\text{TP}}{2\text{TP} + \text{FP} + \text{FN}}$}.
The configurations of the NN and NGBoost were determined based on preliminary experimentation and design considerations. 

\begin{table}[h!]
\caption{Summary of the optimised score thresholds used to determine each classifier's positive and negative predictions based on \cref{eqn:prediction}.}\label{tab:score threshold}
\centering
\begin{tabular}{ccc}
\hline \hline
Classifier & Optimised score threshold  \\
KDE & 0.500\\
KNN & 0.175\\
SVM & 0.560\\
RF & 0.368\\
NN & 0.479\\
NGBoost & 0.128\\
GMM & 0.469\\
XGBoost & 0.497\\
\hline
\end{tabular}
\end{table}
For our custom NN, we employed a focal loss function to address class imbalance. To further help with this challenge, we generated an additional 100 NSWD samples for each NSWD binary system by sampling from the uncertainties derived from the Fisher information matrix. 
Each NSWD data point was weighted inversely proportional to the number of occurrences of its source system to reduce bias introduced by multiple harmonics originating from the same binary. 
The network architecture consists of an input layer, six hidden layers, and an output layer.
ReLU activations are used between layers, and 30\% dropout was applied in the first three hidden layers and subsequently 20\% dropout was applied to the last three hidden layers the network. 
The final output layer employs a sigmoid activation. Training was performed for 100 epochs with a batch size of 128 for both training and inference.  

For NGBoost, we used the default implementation without additional modifications. Lastly, for our custom GMM, we followed the approach described in \cite{Bailer_Jones_2019}. We incorporated custom class priors based on the relative abundance of WDWD and NSWD systems in the Milky Way, set to $\{0: 0.96, 1: 0.04\}$. Since the GMM classifier already incorporates prior information in its probability estimates, no additional probability calibration was applied.

The optimised decision threshold values, as described in \cref{sec:perf_binary}, which set the required predicted probabilities before classifying a system as NSWD, are shown in \cref{tab:score threshold}.

\subsection{Bayesian Optimisation} \label{sec:bayes_opt}
All of our classifiers have one or more hyper-parameters, which are configuration variables that govern the structure and learning behaviour of a model, but are not directly learned from the training data. Examples include the number of trees in a RF, the learning rate in gradient boosting algorithms, or the kernel type in an SVM.
Traditional hyper-parameter optimisation often involves exhaustive grid searches,  which systematically evaluate all possible combinations within a predefined parameter space. However, this is often inefficient and random searches could perform better than grid searches, especially when the dimensionality and complexity increases \citep{JMLR:v13:bergstra12a}. 

In contrast, Bayesian Optimisation \citep{garnett2023bayesian} frequently produces better results with fewer evaluations \citep{BayesianOptimization2012, TakingHuman2016}. Its usage and popularity have risen considerably in recent years, to the extent that widely adopted libraries such as \texttt{scikit-learn} now provide dedicated functions for Bayesian Optimisation \citep{scikit-learn}, which is what we used for our optimisation process.

\subsection{Bayes' Rule Probability Calibration}\label{sec:bayesrule}
The predicted probability from classifiers is not necessarily calibrated to represent the actual probability of the underlying data. In practice, it is often an algorithmic confidence rather than a Bayesian probability \cite[see][for an example on RF classifiers]{Berbel_2024}. Therefore, we applied probability calibration. A traditional probability calibrator learns a direct mapping from model scores to calibrated probabilities using a held-out dataset. This approach implicitly assumes that the observed data are noise-free, and all uncertainty lies in the model output, which is reduced through calibration \citep{muehleisen2016bayesian}. In addition, when class imbalance is severe, as in our case, the probability will be heavily skewed towards 0 or 1, even after calibrating.
Hence, we adapted the calibration method described in \cite{Berbel_2024} to obtain probabilities that both incorporate uncertainty and remain well-defined. This method uses Bayes' rule to calibrate the classifier's output into an informative probability.

We begin by defining $S(\mathbf{X})$ as the distribution of classifier score outputs, where $\mathbf{X}$ denotes the feature inputs from a held-out dataset. 
In practice, this corresponds to the \texttt{predict\_proba} function of most \texttt{scikit-learn} classifiers. To model the distribution of scores conditional on the true label, we split the outputs into two groups: those from the positive class $S(\mathbf{X} \mid y = 1)$  and those from the negative class $S(\mathbf{X} \mid y = 0)$. By fitting a kernel density estimator we obtain smooth approximations of the class-conditional probabilities: $p(S \mid y = 1)$ and $p(S \mid y = 0)$. These densities are further stabilised using a Savitzky–Golay smoothing filter to reduce noise \citep{SV-filter}. We also compute empirical class priors, $\pi_1 = P(y=1 \mid \mathbf{X})$ and $\pi_0 = P(y=0 \mid \mathbf{X})$, essentially assuming the priors to be the distribution of classes in the held-out dataset. Lastly, applying Bayes' Theorem for class 1:

\begin{equation}
    P( y = 1 \mid S)  = \frac{p(S \mid y = 1) \pi_1}{p(S \mid y = 1) \pi_1 + p(S \mid y = 0) \pi_0},
\end{equation}

 and similarly can be achieved for $P( y = 0 \mid S)$. This posterior is computed on a grid of score values in $[0,1]$, yielding a calibrated mapping from uncalibrated classifier scores to probabilities. Finally, we interpolate this mapping to construct a continuous calibration function, which can then be applied to unseen test-set scores to obtain calibrated probabilities.

\section{Hyper-parameter Tuning for XGBoost}
\label{app: hyper}
The results obtained for the hyper-parameters after using Bayesian optimisation for XGBoost classifier in the multi-class classification case can be found in \cref{tab:hyperparameter_multiclass_xgboost} and the binary classification case can be found in \cref{tab:hyperparameter_binaryclass_xgboost}.
\begin{table}[h!]
\caption{Summary of sampled hyper-parameters obtained from Bayesian optimisation for the multi-class XGBoost classifier and their assumed priors. 
$\mathcal{U}(a,b)$ denotes a uniform distribution between $a$ and $b$, while $\mathcal{U}_{\log}(a,b)$ denotes a log-uniform distribution between $a$ and $b$, i.e. $\log x \sim \mathcal{U}(\log a, \log b)$. Parameters not listed in the table were fixed to their default values.}\label{tab:hyperparameter_multiclass_xgboost}
\centering
\begin{tabular}{ccc}
\hline \hline
Hyper-parameter & Prior & Value \\
\hline
\texttt{colsample\_bytree} & $\mathcal{U}(0.5, 1)$ & 0.61317 \\
\texttt{learning\_rate} & $\mathcal{U}_{\log}(10^{-3}, 0.3)$ & 0.052226 \\
\texttt{n\_estimators} & $\mathcal{U}(50, 1000)$ & 701 \\
\texttt{max\_depth} & $\mathcal{U}(3,12)$ & 10 \\
\hline
\end{tabular}
\end{table}

We briefly summarize the tuned parameters here.
The parameter \texttt{max\_depth} limits the maximum depth of each decision tree, preventing overfitting. \texttt{n\_estimators} defines the number of boosting rounds, while \texttt{learning\_rate} scales the contribution of each tree to improve generalisation. \texttt{colsample\_bytree} and \texttt{subsample} specify the fraction of features and training samples used per tree, introducing randomness to reduce overfitting. \texttt{gamma} sets the minimum loss reduction required for further partitioning, and \texttt{min\_child\_weight} controls the minimum sum of instance weights in a child node, acting as another form of regularisation.

For further reading, a detailed description of all available parameters and their tuning guidelines can be found in the official XGBoost documentation:
\url{https://xgboost.readthedocs.io/en/stable/parameter.html}

\begin{table}
\caption{As \cref{tab:hyperparameter_multiclass_xgboost} but for the binary XGBoost classifier for the low-mass population.
}\label{tab:hyperparameter_binaryclass_xgboost}
\centering
\begin{tabular}{ccc}
\hline \hline
Hyper-parameter & Prior & Value \\
\hline
\texttt{colsample\_bytree} & $\mathcal{U}(0.5, 1)$ & 0.85172 \\
\texttt{learning\_rate} & $\mathcal{U}_{\log}(10^{-3}, 0.3)$ & 0.0019931 \\
\texttt{n\_estimators} & $\mathcal{U}(50, 3000)$ & 2626 \\
\texttt{max\_depth} & $\mathcal{U}(3,24)$ & 11 \\
\texttt{gamma} & $\mathcal{U}(0,5)$ & 2.0579 \\
\texttt{min\_child\_weight} & $\mathcal{U}(1,10)$ & 5 \\
\texttt{subsample} & $\mathcal{U}(0.5,1)$ & 0.99617 \\
\hline
\end{tabular}
\end{table}

\end{appendix}

\end{document}